\documentclass[12pt,letterpaper]{JHEP3}         
\usepackage{amssymb,amsfonts}

\pdfoutput=1

\usepackage{epsfig}

\def\be{\begin{eqnarray}}
\def\ee{\end{eqnarray}}
\newcommand{\nn}{\nonumber}
\newcommand\para{\paragraph{}}

\newcommand{\eqn}[1]{(\ref{#1})}

\def\Dslash{\,\,{\raise.15ex\hbox{/}\mkern-12mu D}}
\def\Dbarslash{\,\,{\raise.15ex\hbox{/}\mkern-12mu {\bar D}}}
\def\delslash{\,\,{\raise.15ex\hbox{/}\mkern-9mu \partial}}
\def\delbarslash{\,\,{\raise.15ex\hbox{/}\mkern-9mu {\bar\partial}}}
\def\pslash{\,\,{\raise.15ex\hbox{/}\mkern-9mu p}}
\def\calDslash{\,\,{\raise.15ex\hbox{/}\mkern-12mu {\cal D}}}

\newcommand{\Tr}{{\rm Tr}}

\def\lae{\mathrel{\mathop{\smash{\lower .5 ex \hbox{$\stackrel<\sim$}}}}}
\def\lae{\mathrel{\mathop{\smash{\lower .5 ex \hbox{$\stackrel>\sim$}}}}}

\title{Monopoles and Wilson Lines}

\author{David Tong and Kenny Wong\\
Department of Applied Mathematics and Theoretical Physics, \\
University of Cambridge, \\
Cambridge, CB3 0WA, UK\\{\tt d.tong, k.wong@damtp.cam.ac.uk}}

\abstract{We present a semi-classical description of BPS monopoles interacting with Wilson lines. The Wilson lines are represented as non-Abelian spin impurities. These spins interact  with the monopole degrees of freedom through a natural connection on the moduli space. 
We employ this technology in ${\cal N}=2$ $SU(2)$ gauge theory to count the number of framed BPS states of a single monopole bound to Wilson lines in different representations. 
}

\begin{document}
\pagestyle{plain} \setcounter{page}{1}
\newcounter{bean}
\baselineskip16pt \setcounter{section}{0}

\section{Introduction and Results}

Magnetic monopoles are interesting. At weak coupling, they arise as solitons in non-Abelian gauge theories \cite{thooft,polyakov}. Tracing the fate of these monopoles to the strong coupling regime has proven to be a fruitful technique to understand the dynamics, phase structure and duality properties of quantum field theories.

\para
The  low-energy dynamics of BPS monopoles is best described using the  moduli space approximation \cite{manton}.  The moduli space ${\cal M}$ is the space of solutions to the classical monopole equations; it can be thought of as the configuration space of  monopoles. The low-energy dynamics is governed  by a sigma-model with target space  ${\cal M}$, endowed with a natural metric that is induced by the underlying gauge theory. This means that the classical  scattering of monopoles is described by geodesic motion on ${\cal M}$ while quantum states correspond to functions or forms over ${\cal M}$.

\para
The purpose of this paper is to extend the moduli space description of monopole dynamics to  situations where there are also external quarks, sitting at fixed positions in space. Such fixed quarks are usually represented by the insertion of a Wilson line operator in the path integral,
\be W_R[A_0]={\rm Tr}_R \ {\cal P} \exp\left(i\int dt\,A_0\right)\label{theline}\ee
Here $R$ denotes some representation of the gauge group which, for the purposes of this paper, we take to be $SU(N)$. 
Wilson lines, and their supersymmetric generalisations, are crucial to our understanding of quantum field theory, from their original role as order parameters for confinement \cite{wilson}, to recent discussions, elucidating more subtle  aspects of gauge symmetry and supersymmetry \cite{framed,line}. 

\para
In order to describe the way monopoles interact with external quarks, it will prove useful to work with a different description of the Wilson line.  The starting point is a simple, semi-classical model of an external quark, viewed as a fixed electric source for the non-Abelian gauge field. Such a quark carries colour degrees of freedom, described by a {\it spin}, or vector of fixed length, and the background gauge field causes this spin to precess.  It is straightforward to show  that the quantum mechanical path integral for such a spin is equal to the Wilson line: $Z_{\rm spin}[A_0]=W_R[A_0]$.
In other words, the Wilson line can be thought of as the effect of integrating out localised spin degrees of freedom. We review this perspective on the Wilson line in Section \ref{wilsec}. 

\para
The main result in this paper is contained in Section \ref{monosec} where we explain how magnetic monopoles couple to the localised spins and, through this, to Wilson lines. As we shall see, this too has an elegant geometrical description in the moduli space language. While the underlying gauge dynamics induce a natural metric on ${\cal M}$, it also provides the moduli space with a number of further geometric quantities. Among these is an $SU(N)$ gauge connection. We will show that the electric degrees of freedom of the external quark couple to this connection. As we will see, this leads to the expected non-Abelian Lorentz-force law for the centre of mass motion of monopoles in the presence of an electric charge. But it also leads to more subtle dynamics in which moving monopoles exchange electric charge with fixed, external quarks. 

%
%

\para
Finally, in Section \ref{boundsec}, we use this approach to study  the supersymmetric quantum mechanics of monopoles in ${\cal N}=2$ $SU(2)$ Yang-Mills. We compute some simple examples of  framed BPS states \cite{framed}, involving quantum monopoles bound to Wilson lines in different representations.

\section{Spin Impurities and Wilson Lines}\label{wilsec}

In this first section, we explain how spin impurities, coupled to bulk gauge fields, can be thought of as Wilson lines.  The essence of these ideas is not new. Early discussions were given, for example, in \cite{bala,dp}. Mathematically, the relationship uses the framework of geometric quantization and a  readable exposition can be found in \cite{beasley}. Here, however,  we
take a different tack. Our goal is to describe the connection between Wilson lines and spin impurities in a pedestrian manner without ever mentioning the words ``nilpotent orbit".

\subsection{Classical Spin}

Classically, we view a spin as an $N$-component complex vector $w_a$, $a=1,\ldots,N$ of fixed length,
\be w^\dagger w = \kappa\label{ww}\ee
We further identify vectors which differ only by a phase: $\omega_a\sim e^{i\theta} w_a$. This means that the vectors parameterise the projective space ${\bf CP}^{N-1}$.

\para
To implement the phase equivalence of vectors, we  introduce an auxiliary $U(1)$ gauge field $\alpha$ which lives on the worldline of the spin. The action is 
\be S  = \int dt\  \left(iw^\dagger{\cal D}_t  w - \kappa\alpha\right)\nn\ee
where ${\cal D}_t = \partial_t w - i \alpha w$. There is now a gauge symmetry $w\rightarrow e^{i\theta(t)}w$. Correspondingly,  the gauge field acts as a Lagrange multiplier, implementing the constraint \eqn{ww}. Note that $\kappa$ appears as a one-dimensional Chern-Simons term in the action and we will see below that there is an associated quantisation condition on $\kappa$. 

\para
Importantly,  because the action is first order, rather than second order, the classical spin has a phase space, rather than configuration space,  given by ${\bf CP}^{N-1}$. As we will see shortly, upon quantisation this compact phase space results in a finite dimensional Hilbert space. Actions of this kind are familiar from the discussion of classical spin-1/2 particles \cite{stone}. 

\para
So far our spin has no dynamics. This arises by coupling it to a background $SU(N)$ gauge field $A_\mu$. This gauge field propagates in $d+1$ spacetime dimensions and is governed by its own equations of motion which we shall consider later in the paper. Meanwhile, the spin impurity sits at a fixed position in space which we will take to be the origin. The action for the spin is 
\be S_{\rm spin} = \int dt\ \left(iw^\dagger{\cal D}_t  w - \kappa\alpha   -w^\dagger A_0(t) w \right)\label{spinact}\ee
where $A_0(t) = A_0(\vec{x}=0,t)$ is the value of the temporal gauge field at the origin. Physically, we can think of this spin impurity as a classical quark. Although the position of the quark is fixed at the origin, its gauge orientation is described by $w_a$ and is free to fluctuate. The classical equations of motion now tell us that the background gauge field causes the spin to precess. In $\alpha=0$ gauge, we have
\be i\frac{dw}{dt} = A_0(t) w\nn\ee
This has solution 
\be w(t) = {\cal P} \exp\left( i\int^t_{t_0} dt'\ A_0(t')\right)w (t')\nn\ee
where ${\cal P}$ stands for path ordering. We see that, already classically, the unitary operator associated to the Wilson line plays a role in the dynamics of this system. However, our real interest is in the quantum story.

\subsection{Quantum Spin}

It is a simple matter to quantise the spin system. It is easiest to first work in the Hamiltonian formalism, starting with the unconstrained variables, $w_a$. These obey the commutation relations,
\be [w_a,w_b^\dagger] = \delta_{ab}\nn\ee
We define a ``ground state" $|0\rangle$ such that $w_a|0\rangle=0$ for all $a=1,\ldots,N$. A general state in the Hilbert space then takes the form
\be |a_1\ldots a_n\rangle = w_{a_1}^\dagger\ldots w_{a_n}^\dagger|0\rangle\nn\ee
In the quantum theory, there is a normal ordering ambiguity in defining the  constraint \eqn{ww}. The symmetric choice is to take the charge operator
\be Q =\frac{1}{2} (w_a^\dagger w_a + w_a w_a^\dagger)\label{qsym}\ee
and to impose the constraint
\be Q = \kappa\label{qk}\ee
The spectrum of $Q$ is quantised which means that the theory only make sense if the Chern-Simons coefficient $\kappa$ is also quantised. 
However, the normal ordering implicit in the symmetric choice of  $Q$ in \eqn{qsym} gives rise to a shift in the spectrum. For $N$ even, $Q$ takes integer values; for $N$ odd, $Q$ takes half-integer values. It will prove useful to introduce the shifted Chern-Simons coefficient, 
\be \kappa_{\rm eff} = \kappa - \frac{N}{2}\nn\ee
The promised quantisation condition then reads $\kappa_{\rm eff} \in {\bf Z}^+$.

\para
The  constraint \eqn{qk} restricts the theory to a finite dimensional Hilbert space,  as expected from the quantisation of a compact phase space ${\bf CP}^{N-1}$. Moreover, for each value of $\kappa_{\rm eff}$, the Hilbert space inherits an action under the $SU(N)$ global symmetry. Let us look at some examples:
\begin{itemize}
\item $\kappa_{\rm eff} = 0$: The Hilbert space consists of a single state, $|0\rangle$.
\item $\kappa_{\rm eff} = 1$: The Hilbert space consists of $N$ states, $w_a^\dagger |0\rangle$, transforming in the fundamental representation of $SU(N)$.
\item $\kappa_{\rm eff} = 2$: The Hilbert space consists of $\frac{1}{2}N(N+1)$ states, $w_a^\dagger w_b^\dagger |0\rangle$, transforming in the symmetric representation. 
\end{itemize}
By increasing the value of $\kappa_{\rm eff}$ in integer amounts, it is clear that we can build all symmetric representations of $SU(N)$ in this manner.

\para
The states in the Hilbert space can be interpreted as the lowest Landau level states on ${\bf CP}^{N-1}$ with $\kappa_{\rm eff}$ units of magnetic flux, which are known to transform in the symmetric product of $\kappa_{\rm eff}$  copies of the fundamental representation of $SU(N)$ \cite{nair}.

\subsection{The Path Integral}

Let us now consider the  path integral formulation of a quantum spin.  From our discussion above, we expect that the path integral will be  non-vanishing only when evaluated on some finite-dimensional Hilbert space of states determined by $\kappa$. Anticipating this, we will 
insert $p$ creation operators at $t=-\infty$ and a further $p$ annihilation operators at $t=+\infty$ in the path integral and compute
\be  Z_{\rm spin}[A_0] = \frac{1}{p!}\int {\cal D}\alpha{\cal D}w{\cal D}w^\dagger\,e^{iS_{\rm spin}}\, w_{a_1}(+\infty)\ldots w_{a_p}(+\infty)\,w^\dagger_{a_1}(-\infty)\ldots w^\dagger_{a_p}(-\infty)\nn\ee
To evaluate this partition function,  we work with the propagator $\theta(t_1-t_2)\delta_{ab}$ for the field $w_a$, where $\theta$ is the Heaviside step function.

\para We deal first with the vacuum bubbles. They exponentiate, evaluating to 
%
%
%
\be
&&\prod_{n = 1}^\infty  \exp \Bigg( \frac{i^n}{n} \int dt_1 ... dt_n (A_0(t_1)_{a_1a_2}+\alpha(t_1)\delta_{a_1a_2}) \theta(t_1 - t_2) \nn\\ &&\ \ \ \ \ \ \ \  \times(A_0(t_2)_{a_2 a_3} +\alpha(t_2)\delta_{a_2a_3})\theta(t_2 - t_3)  \ldots (A_0(t_n)_{a_n a_1} +\alpha(t_n)\delta_{a_na_1})\theta(t_n - t_1) \Bigg)\nn
\ee
All $n \geq 2 $ factors vanish because the product of the step functions vanishes everywhere except on a set of measure zero. We're left only with the $n=1$ factor. This is independent of the background $SU(N)$ gauge field $A_0$ because it is traceless. However, it does depend on $\alpha$. Using the midpoint regularisation $\theta(0)=\frac{1}{2}$, the net effect of these vacuum bubbles is to renormalise the Chern-Simons term $\kappa\rightarrow \kappa-N/2 = \kappa_{\rm eff}$. (This well known result was first derived in \cite{elitzur}). The 
 path integral is invariant  under large gauge transformations only if $\kappa_{\rm eff}\in {\bf Z}$. This reproduces the quantisation condition we saw using the Hamiltonian approach above\footnote{There are a number of small, and ultimately unimportant, differences between our calculation and the standard calculation presented in \cite{elitzur} and reviewed in Section 5.5. of \cite{dunne}. First, the calculation was done in these papers for fermions, but the result for bosons with first order kinetic terms differs only by a minus sign. Second, our shift of the Chern-Simons term does not depend on the sign of the ``mass" of the boson which, for us, translates in the sign of the eigenvalues of $A_0$. This can be traced to our choice of propagator for all fields $w_a$ regardless of their mass.  This is the  appropriate choice to agree with the  vacuum state $w_a|0\rangle=0$ that we employed in the Hamiltonian quantisation.}.

\subsubsection*{Fundamental Representation}

\begin{figure}[!h]
\begin{center}
\includegraphics[ width=6.2in, height = 0.75in]{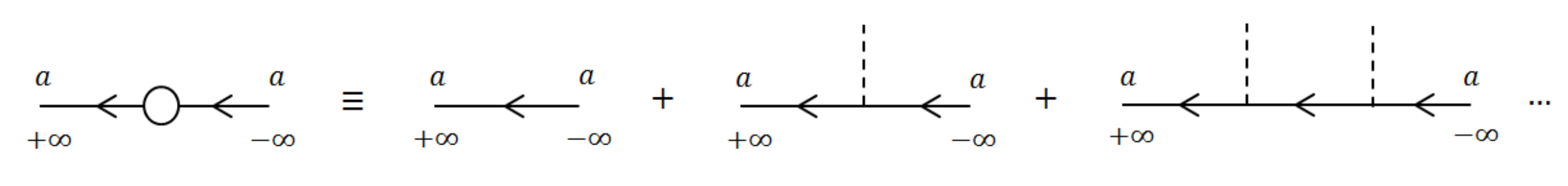}
\end{center}
\caption{The $p=1$ diagrams}
\label{feynman1}
\end{figure}

Let's now complete the evaluation of the path integral. For $p=1$, we have just two insertions in the path integral
\be
Z_{\rm spin}[A_0] = \int {\cal D}\alpha{\cal D}w{\cal D}w^\dagger\,e^{iS_{\rm spin}}\, w_{a}(+\infty) w^\dagger_{a}(-\infty)\nn\ee
We first do the path integral over $w$ and $w^\dagger$. Having summed the vacuum bubbles above, we're left with the series of diagrams shown in Figure \ref{feynman1}.
These diagrams correspond to the sum
\be
&& \delta_{aa} + i \int dt_1 (A_0(t_1)_{aa}+\alpha(t_1)\delta_{aa}) \nn \\
&& \ \ \  \  \  \ \  \ - \int dt_1 dt_2  (A _0(t_1)_{ab}+\alpha(t_1)\delta_{ab} )\theta (t_1 - t_2) (A(t_2)_{ba} +\alpha(t_2)\delta_{ba}) -\ldots \nn \ee
 These can  be easily summed to give the time ordered trace which, up to a phase, we recognise as the Wilson line \eqn{theline}, 
 \be
 {\cal P} \exp \left(i \int dt\ (A_0(t)+\alpha(t)) \right)_{aa} = W[A_0]\,e^{i\int dt \, \alpha(t)}\nn
\ee
Here the Wilson line is evaluated in the fundamental representation. Putting this together with the vacuum bubbles, we're left with the partition function
\be 
Z_{\rm spin}[A_0] &=&  W[A_0]\int {\cal D}\alpha \,e^{-i\int dt (\kappa_{\rm eff}-1)\alpha(t) }\nn\ee
The remaining integral over $\alpha$ is simple: it acts as a delta function, giving a non-vanishing answer only if $\kappa_{\rm eff}=1$. But this is what we expect from our discussion of the Hamiltonian quantisation: only when $\kappa_{\rm eff} = 1$ is the Hilbert space $N$-dimensional with an fundamental action of $SU(N)$. In this case, we have simply
\be Z_{\rm spin}[A_0]=W[A_0]\nn\ee
This is the result that we wanted: the partition function of the spin impurity is the Wilson line for the $SU(N)$ gauge field, here evaluated in the fundamental representation.

\subsubsection*{Symmetric Representations}

It is simple to extend the discussion above to higher symmetric representations by considering $p\geq 2$ insertions. For example, for $p=2$ the diagrams are
\begin{figure}[!h]
\begin{center}
\includegraphics[ width=3.2in, height = 1.1in]{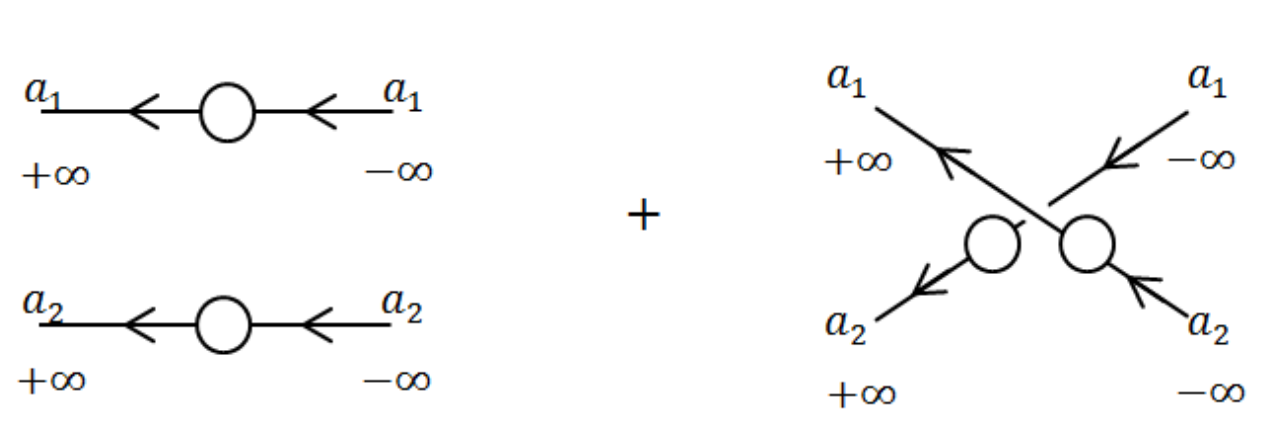}
\end{center}
\caption{The $p = 2$ diagrams}
\label{feynman2}
\end{figure}

\noindent
The path integral now factorises into the symmetric product, 
\be
Z_{\rm spin}[A_0] = \delta_{k_{\rm eff},p}\,\frac 1 {p!} \sum_{\sigma \in S_p} {\cal P} \exp \left(i \int dt \,A_0(t) \right)_{a_1 a_{\sigma(1)}} \times ... \times \mathcal P \exp \left( i\int dt A_0(t) \right)_{a_p a_{\sigma(p)}} \nn\ee
where, as before, the overall delta-function arises from the integral over $\alpha$ and requires $\kappa_{\rm eff} =p$. This is now the Wilson line, 
\be Z_{\rm spin}[A_0] = W_R[A_0]\nn\ee
with $R$ the $p^{\rm th}$ symmetric tensor product of the fundamental representation. A very similar construction of Wilson loops in the symmetric representation using D-branes was given in \cite{gomis}.

\subsubsection*{Anti-Symmetric Representations}

We can construct Wilson lines in the anti-symmetric representation by retaining the action \eqn{spinact}, but quantising the spin degrees of freedom as fermions, with anti-commutation relations
\be \{w_a,w_b^\dagger\} = \delta_{ab}\nn\ee
The discussion above goes through essentially unchanged apart from a few minus signs. For our purposes, the most important of these is the relative minus sign between the two diagrams in Figure \ref{feynman2}. The end result is that the partition function with $p$ insertions now computes the Wilson line in the $p^{\rm th}$ anti-symmetric representation.

\section{Monopole Dynamics}\label{monosec}

We now turn to our main story: the interaction of  monopoles with Wilson lines.    A moduli space description of this dynamics was previously developed in \cite{sungjay} but focusses only on Abelian long-range fields. (This approximation is valid near walls of marginal stability which was the main interest in that paper). Here, instead, we treat the full non-Abelian dynamics of both the monopole and Wilson line. 

\para
We start by reviewing the monopole solutions and their moduli space dynamics in the absence of impurities. More detailed discussions can be found in any number of reviews such as \cite{harvey,nick,tasi}.

\subsection{A Review of Monopoles}

Throughout this section, we work with $SU(2)$ Yang-Mills theory. (The extension to monopoles in higher rank gauge groups is straightforward). The gauge potential $A_\mu$ is accompanied by a pair of real, adjoint scalar fields that we call $\phi$ and $\sigma$. The action is 
\be
S_{YM} = \frac{1}{e^2} \int d^4 x \ {\rm Tr}\left( - \frac{1}{2} F_{\mu\nu} F^{\mu\nu} - {\cal D}_\mu\phi {\cal D}^\mu\phi-{\cal D}_\mu \sigma {\cal D}^\mu \sigma  + [\phi,\sigma]^2\right)
\label{4dact}\ee
This can be viewed as part of an action with either ${\cal N}=2$ or ${\cal N}=4$ supersymmetry. 

\para
We are interested in the phase of the theory where $SU(2)$ gauge symmetry is broken down to $U(1)$ by a vacuum expectation value,
\be \langle \Tr\,\phi^2\rangle =\frac{v^2}{2}\nn\ee
We will chose the expectation value for the other scalar field to be $\langle \sigma\rangle =0$. Indeed, it will not play a role until we introduce the Wilson line in Section \ref{monlinesec}.

\para
This theory famously admits magnetic monopole soliton solutions \cite{thooft,polyakov}. They can be obtained by setting $\sigma=0$, while the remaining fields obey the Bogomolnyi equation, 
\be B_i = {\cal D}_i\phi\label{bog}\ee
where $i=1,2,3$ label spatial indices and the non-Abelian magnetic field is defined by $B_i=\frac{1}{2}\epsilon_{ijk}F_{jk}$.

\para
The magnetic charge of the solution is determined by the  topological winding number $n\in {\bf Z}$ of the field $\phi$ at spatial infinity.  Solutions to these equations have mass
\be M_{\rm mono} = \frac{4\pi v |n|}{e^2}\label{mmass}\ee
The linearity of this mass formula suggests that there is no classical force between $n$ separated monopoles. This intuition is borne out by index theorems which show that the general solution has $4n$ collective coordinates \cite{erick}.  For far-separated monopoles, these can be thought of as the positions of $n$ charge-one monopoles moving in ${\bf R}^3$, each of which carries an extra internal degree of freedom. In contrast, as the monopoles approach each other, they lose their individual identities and the interpretation of the collective coordinates becomes more complicated. 

\para
We write the most general solution as $A_i(x;X^\alpha)$ and $\phi(x;X^\alpha)$ where $X^\alpha$, $\alpha = 1,\ldots, 4N$, are collective coordinates. These parameterise the monopole moduli space which takes the form
\be {\cal M}_n \cong {\bf R}^3\times \frac{{\bf S}^1\times \tilde{\cal M}_n}{{\bf Z}_n}\label{mspace}\ee
Here the ${\bf R}^3$ factor parameterises the centre of mass motion of the monopoles while the ${\bf S}^1$ factor arises from large gauge transformations of the unbroken $U(1)$ gauge group. The $4(n-1)$ dimensional manifold $\tilde{M}_n$ describes the relative positions and internal phases of the magnetic monopoles. 

\subsection{Moduli Space Dynamics}\label{mdynamicssec}

The dynamics of slowly moving magnetic monopoles is well captured by the {\it moduli space approximation}. Heuristically, the idea is that if we were to take a snapshot of the field configuration at any time then it would look close to a static configuration labelled by a point in ${\cal M}_n$. This means that we can reduce a field theoretic problem to a much simpler problem of dynamics on the moduli space ${\cal M}_n$.

\para
To describe the moduli space dynamics in more detail, we start by introducing a zero mode associated to each collective coordinate. 
The zero mode is defined as the derivative of each field, together with an accompanying gauge transformation,
\be \delta_\alpha A_i = \frac{\partial A_i}{\partial X^\alpha} - {\cal D}_i\Omega_\alpha\ \ \ ,\ \ \ \ \delta_\alpha \phi = \frac{\partial \phi}{\partial X^\alpha} + i[\phi,\Omega_\alpha]\nn\ee
By construction, the zero mode is a solution to the linearised Bogomolnyi equation.  The gauge transformations $\Omega_\alpha(x,X)$ will be important in what follows. They are designed to solve the background gauge fixing condition,
\be {\cal D}_i \,(\delta_\alpha A_i) - i [\phi,\delta_\alpha\phi]=0\label{bgauge}\ee
With these zero modes in hand, we can describe the low-energy dynamics of monopoles \cite{manton}. To this end, we promote the collective coordinates $X^\alpha$ to time-dependent degrees of freedom, $X^\alpha(t)$. When  the monopoles move they generate a non-Abelian electric field $E_i=F_{0i}$ given by
\be E_i = \frac{\partial A_i}{\partial X^\alpha}\dot{X}^\alpha - {\cal D}_iA_0\nn\ee
where $A_0$ must be chosen so that Gauss' law is satisfied:
\be {\cal D}_i E_i - i[\phi,{\cal D}_0\phi]=0\nn\ee
This is achieved by
\be A_0 = \Omega_\alpha(x;X) \dot{X}^\alpha\label{a01}\ee
which ensures that Gauss' law is obeyed by virtue of the gauge fixing condition \eqn{bgauge}.  Substituting these expressions into the original action \eqn{4dact}, we arrive at an expression for the dynamics of monopoles which takes the form of a sigma-model on the moduli space ${\cal M}_n$,
\be S_{\rm mono} = \int dt\ \frac{1}{2}\,g_{\alpha\beta}(X)\,\dot{X}^\alpha\dot{X}^\beta\label{monact}\ee
where the metric is given by the overlap of zero modes
\be
g_{\alpha \beta} (X) = \frac{2}{e^2} \int d^4 x \ {\rm Tr}\, \left( \delta_\alpha A_i \, \delta_\beta A_i  +\delta_\alpha\phi\,\delta_\beta\phi\right)
\label{mmetric}\ee
This metric has a number of special properties. It is geodesically complete, hyperK\"ahler and inherits an $SU(2)\times U(1)$ isometry from spatial rotations and global gauge transformations of the underlying field theory. For a single monopole, it is simply the flat metric on ${\cal M}_1\cong {\bf R}^3\times{\bf S}^1$. For a pair of monopoles, the metric on $\tilde{\cal M}_2$ has been explicitly constructed and is known as the Atiyah-Hitchin metric \cite{ah}. For $n>2$ monopoles, the metric is known only asymptotically \cite{gm}.

\subsubsection*{The Connection}

We will shortly introduce Wilson lines into the game and see how they affect the dynamics of monopoles. But we have already met the most important ingredient needed for this discussion: the connection $\Omega_\alpha(x;X)$. We pause here to explain why this can be thought of as an $SU(2)$ gauge connection over the moduli space ${\cal M}_n$. 

\para
Let us first show that it transforms as a connection. Suppose that we chose to present the monopole solutions in a different gauge by acting with a gauge transformation $g(x;X)\in SU(2)$. This means that
\be A'_i = gA_ig^{-1} + ig\partial_i g^{-1}\ \ \ \ ,\ \ \ \ \phi'= g\phi g^{-1}\nn\ee
Our goal is to find the new compensating gauge transformation $\Omega'_\alpha(x;X)$ such that the zero modes $\delta_\alpha A'_i=\partial_\alpha A'_i - {\cal D}'_i\Omega'_\alpha$ and $\delta_\alpha \phi' = \partial_\alpha \phi' - i[\phi',\Omega'_\alpha]$ obey the background gauge condition ${\cal D}'_i\,\delta_\alpha A'_i - i[\phi',\delta\phi']=0$. A short calculation shows that this is satisfied if we chose
\be \Omega'_\alpha = g\Omega_\alpha g^{-1} +i g\partial_\alpha g^{-1}\nn\ee
 This is the statement that $\Omega_\alpha$ transforms as a gauge connection over ${\cal M}_n$.

\para
There remains a small issue.  The connection $\Omega_\alpha(x;X)$ appears to depend on both spatial coordinates $x$ and moduli space coordinates $X$. This is misleading. Because the ${\bf R}^3$ factor of the moduli space \eqn{mspace} describes the centre of mass motion of the monopoles, the gauge transformations always take the form $\Omega_\alpha(x-X,\tilde{X})$ where $\tilde{X}$ parameterise the remaining ${\bf S}^1$ and $\tilde{\cal M}_n$ factors. This means that we can always restrict attention to the point  $x=0$ without losing information. It is the   resulting object,  $\Omega_\alpha(x=0,X)$, which acts as an $SU(2)$  connection over ${\cal M}_n$.

\subsection{Monopoles and Wilson Lines}\label{monlinesec}

It is now time to introduce Wilson lines into our monopole story.  We choose to place the  Wilson line at the spatial origin, $x^i=0$. Following our discussion in Section \ref{wilsec}, we represent the Wilson line by introducing spin impurities $w_a$, with $a=1,2$, subject to the requirement that $w^\dagger w= \kappa$. The action for these spins is
\be
S = S_{YM} + \int d^4x \ \left[ i w^\dagger {\cal D}_t w -\kappa \alpha + w^\dagger (A_0-\sigma) w\right]\delta^3(x)\label{spinimp}
\ee
One difference from the discussion in Section \ref{wilsec} is that the spin impurities couple to both the gauge field and the scalar field $\sigma$.  This ensures that the spin impurities are 1/2-BPS. Performing the path integral over the spin degrees of freedom leaves us with the Yang-Mills partition function with an insertion of 
\be W_R = {\rm Tr}_R\,{\cal P}\exp\left( i\int dt\ A_0(t)-\sigma(t)\right)\label{swilson}\ee
with the representation $R$ determined, as in Section \ref{wilsec}, by the Chern-Simons coefficient $\kappa$. This is familiar in the supersymmetric context, where BPS Wilson lines necessarily involve both gauge fields and scalars. This was first demonstrated in ${\cal N}=4$ super Yang-Mills in \cite{malda,sjrey} and the different possibilities allowed by supersymmetry were explored in some detail in \cite{zarembo, trancanelli}\footnote{Viewed as the insertion of electric impurities, the need to couple to an extra scalar field to preserve supersymmetry was noticed in Abelian theories in \cite{shamit} and further explored in \cite{shamit2}. Indeed, the discussion above makes clear that the analog of doping with electric impurities in a non-Abelian gauge theory is  the insertion of (randomly placed) Wilson lines.}.

\para
Here we restrict attention to the simplest, straight Wilson line. (It seems likely that the discussion can be generalised to moving external quarks). The insertion of a spin impurity sources both $A_0$ and $\sigma$. They now obey
\be  -{\cal D}_i E_i + i[\phi,{\cal D}_0\phi] +i[\sigma,{\cal D}_0\sigma] = e^2\,ww^\dagger \,\delta^3(x) \label{e1}\ee
and 
\be {\cal D}^2\sigma - [\phi,[\phi,\sigma]]  = e^2ww^\dagger\,\delta^3(x)\label{sigma1}\ee
For stationary configurations, these are actually the same equation: if we can solve the first, we can solve the second simply by setting
\be A_0=\sigma\label{a02}\ee
Although trivial, this  observation has consequence.  First, because the spin in \eqn{spinimp} couples to $A_0-\sigma$, it means that two  impurities, placed some distance apart, feel neither a repulsive force nor an attractive force between the their spins. The gauge field $A_0$ mediates a repulsive force but is cancelled by the attractive force from $\sigma$. This kind of ``no-force" condition is, of course, almost synonymous with ``BPS". 

\para
Secondly, it means that even though  $A_0$ and $\sigma$ are non-zero, they cancel each other out in the equations of motion for the other fields, $A_i$ and $\phi$. This, in turn, ensures that the solutions to the Bogomolnyi equation \eqn{bog} remain solutions when the spin impurities are inserted. The only difference is that the equations \eqn{e1} and \eqn{sigma1} must now be solved on the background of the monopole solution. The end  result is that the moduli space of monopoles in the presence of a Wilson line is again given by ${\cal M}_n$. 

\para
Our task is to understand the dynamics of the monopoles in the presence of the spin impurities. A very similar problem was recently solved in \cite{us}, studying Abelian vortices in the presence of electric impurities. Within the moduli space approximation, we  again promote the collective coordinates to dynamical degrees of freedom,  $X^\alpha(t)$. However, these now couple to the spin impurities $w^a(t)$, with $a=1,2$. 

\para
As we saw above, when the monopoles move, they  induce an electric field. Our ansatz for $A_0$ is simply a linear combination of \eqn{a01} and \eqn{a02},
\be A_0 = \Omega_\alpha \dot{X}^\alpha + \sigma\nn\ee
With this choice, we have
\be E_i = \delta_\alpha A_i\dot{X}^\alpha - {\cal D}_i\sigma\ \ \ \ , \ \ \ \ \ {\cal D}_0\phi = \delta_\alpha \phi\dot{X}^\alpha + i[\phi,\sigma]\nn\ee
To proceed, we work to leading order in the gauge coupling $e^2$. It is simple to check that $E_i$ and ${\cal D}_0\phi$  obey Gauss' law \eqn{e1} to leading order. However, there are further terms, such as $[\sigma,{\cal D}_0\sigma]$, which are  ${\cal O}(e^4)$ which do not obviously cancel; these can be neglected at the order of our approximation.  A related issue arises when we substitute this ansatz into the action \eqn{spinimp}; the kinetic term $({\cal D}_0\sigma)^2$ are again of order ${\cal O}(e^4)$ and we drop them in what follows\footnote{This same approximation is also necessary in  other contexts where solitons acquire a connection term over their moduli space \cite{us,kimyeong,csvort}. However, considerations of supersymmetry suggest that the final result is nonetheless exact. We expect the same to be true here and this is confirmed by the D-brane picture of solitons interacting with Wilson lines \cite{toappear}.}. 

\para
The remaining kinetic terms $E^2_i$ and $({\cal D}_0\phi)^2$ contribute to the dynamics at leading order. The end result is an action which governs the coupling between the monopoles and spin impurities,
\be S= S_{\rm mono} + \int dt\ \left(iw^\dagger{\cal D}_t  w - \kappa\alpha+ w^\dagger \Omega_\alpha w\,\dot{X}^\alpha\right) \label{final} \ee
where $S_{\rm mono}$ is the sigma-model on the monopole moduli space \eqn{monact} with the usual metric and $\Omega_\alpha = \Omega_\alpha (x^i = 0; X)$. We see that the interaction between the monopoles and spin impurities is mediated by the $SU(2)$ connection $\Omega$ over ${\cal M}$. 

\para
We can derive an alternative description by integrating out the spin degrees of freedom.  In the original Yang-Mills theory, this takes us back to the Wilson line \eqn{theline}. In the effective dynamics of monopoles, we can use the results of Section \ref{wilsec} to derive the monopole partition function,
\be Z_{\rm mono} = \int {\cal D}X\,\widehat{W}_R(X)\,e^{iS_{\rm mono}}\nn\ee
The spin impurities have resulted in the insertion of  $\widehat{W}$, the holonomy of the moduli space connection $\Omega$ along the path $C$ taken in  ${\cal M}_n$,
\be \widehat{W}_R(X) = {\rm Tr}_R \,{\cal P}\,\exp\left( i\int_C \, \Omega_\alpha d X^\alpha \right)\nn\ee
This is our final result for the dynamics of monopoles in the presence of Wilson lines.

\subsubsection*{A Comment on 't Hooft Lines}

Below, we will explore the interactions of monopoles and Wilson lines in more detail. But, first, we make a passing comment. The magnetic dual of Wilson lines are 't Hooft lines. These can be thought of as the insertion of a very heavy, magnetically charged object.

\para
The interaction of monopoles with 't Hooft lines is somewhat different. Both objects can be mutually BPS and solutions exist with the monopole sitting at arbitrary separation from the 't Hooft line. However, in contrast to the Wilson line, the 't Hooft line distorts the monopole solution. This means that the dynamics of monopoles is again described in terms of a sigma-model on the moduli space, but now with a deformed metric \cite{cherkis}. Similar results also hold for $d=2+1$ dimensional vortices moving in the presence of magnetic defects \cite{us}.

\subsection{Classical Scattering} 

Our final expression for the monopole effective action \eqn{final} is defined in terms of various geometric objects over the monopole moduli space. For the case of a single $n=1$ monopole, we now provide more explicit expressions for these objects.

\subsubsection*{A Single Monopole}

For a single monopole, the solution to the Bogomolnyi equation \eqn{bog} is known explicitly. If we place the monopole at the origin, it is
\be
A_i= \left( 1 - \frac{vr}{\sinh vr} \right) \frac{\epsilon^{aij} x_j} {r^2} \frac{\sigma^a} 2, \qquad \varphi = \left(\frac 1 {vr} - \coth vr \right) \frac {vx^a} r\frac{\sigma^a} 2 \label{msol}\ee
The monopole has 4 collective coordinates. Three of these are straightforward: they  correspond to the centre of mass of the monopole. We introduce these translational collective coordinates simply by writing  $A_i=A_i(x-X)$ and $\phi=\phi(x-X)$. The zero modes are then given by
\be 
\delta_i  A_j = \frac{\partial A_j}{\partial X^i} - {\cal D}_j \Omega_i\ \ \ \ ,\ \ \ \ \delta_i\phi = \frac{\partial \phi}{\partial X^i} + i[\Omega_i,\phi]
\nn\ee
where, as explained in Section \ref{mdynamicssec}, the compensating gauge transformation $\Omega_i$ is designed so that the zero modes satisfy the background gauge condition \eqn{bgauge}. For these translational modes, something nice happens: the compensating gauge transformation is given by the gauge connection itself:
\be \Omega_i = -A_i(x - X)\nn\ee
With this choice, the zero modes take the simple form $\delta_iA_j = -F_{ij}$ and $\delta_i\phi= -{\cal D}_i\phi$ and Gauss' law is solved by virtue of the original equations of motion.
 
\para
The fourth collective coordinate, $\chi$, is periodic and  arises from acting on the background \eqn{msol} with large gauge transformations in the unbroken  $U(1)\subset SU(2)$ given by $g=e^{-i\phi \chi}$. The compact nature of the gauge group means that $\chi \in [0,2\pi/v)$. Infinitesimally, this gauge transformation is
\be \Omega_\chi = -\phi(x,X)\nn\ee
which provides the expression for the final piece of the connection on moduli space.  Motion in this $\chi$ direction turns on $A_0=\Omega_\chi \dot{\chi}$ which gives rise to an electric field $F_{0i}= B_i\dot{\chi}$, turning the monopole into a dyon with electric charge 
\be
q_{\rm mono} = {4\pi}\dot{\chi}/e^2\label{mcharge}\ee
The upshot of this discussion is that the moduli space for a single monopole is 
\be {\cal M}_1 \cong {\bf R}^3\times {\bf S}^1\nn\ee
The metric can be computed by taking the overlap of zero modes \eqn{mmetric} and is given simply by
\be ds^2 = M (d\vec{X}^2 + d\chi^2)\nn\ee
where $M = 4\pi v/e^2$ is the mass of a single monopole.

\subsubsection*{The $SU(2)$ Spin}

We can also be more explicit about the spin degree of freedom itself. 
For the $SU(2)$ gauge group, the spin impurity has phase space ${\bf CP}^1$. In this case, it is simplest -- and perhaps more familiar -- to use unconstrained coordinates that parameterise the phase space. In the quantum effective action, the Chern-Simons term $\kappa$ is renormalised to $\kappa_{\rm eff} =\kappa-1$ as explained in Section \ref{wilsec}. We write the two-component spin $w^a$ in polar coordinates as
\be 
w = \sqrt{\kappa_{\rm eff}} e^{i\psi} \left( \begin{array}{c} e^{-i\varphi/2} \cos \frac{\theta}{2} \\  e^{+i\varphi/2} \sin \frac{\theta}{2} \end{array} \right)
\nn\ee
Discarding a total derivative, the kinetic term for the spin in \eqn{spinact} becomes
\be
S_{\rm kin} = \frac{\kappa_{\rm eff}}{2} \int dt\ \dot{\varphi} \cos \theta
\nn\ee
This can be thought of as a Dirac monopole connection for the spin degree of freedom \cite{stone} although, confusingly, one that has nothing to do with the 't Hooft-Polyakov magnetic monopole.  (See, for example, \cite{hitoshi} for a nice pedagogical discussion of the classical and quantum aspects of this simple Lagrangian). 

\para
We can form a triplet of operators that transform in the adjoint of $SU(2)$,
\be
J^a = \frac{1}{2} w^\dagger \sigma^a w = \frac{\kappa_{\rm eff}}{2}\left(\begin{array}{c}\sin\theta\cos\varphi \\ \sin\theta\sin\varphi \\ \cos\theta\end{array}\right)\ \ \ \ \ a=1,2,3\label{ja}\ee
These are analogous to the spin ``angular momentum" operators when discussing representations of the Lorentz group. In the quantum theory, they obey the commutation relations $[J^a,J^b] = i\epsilon^{abc}J^c$. In the present context, $J^a$ determines the electric charge of the spin impurity under the unbroken $U(1)\subset SU(2)$ gauge group,
\be
q_{\rm spin} = \frac{{\rm Tr}( J \phi ) }{v}
\label{spincharge}\ee
where the factor of $v$ ensures that the charge is normalised in the same way as \eqn{mcharge}. Here we have introduced the notation $J=J^a\sigma^a$ so that $J$ lives in the $su(2)$ Lie algebra. For example, when $\kappa_{\rm eff}=1$, the spin lies in the fundamental representation of $SU(2)$ and $q_{\rm spin}\in [-1/2,+1/2]$. When $\kappa_{\rm eff}=2$, the spin lies in the triplet and $q_{\rm spin}\in [-1,+1]$.

\subsubsection*{Scattering}

With these expressions for the monopole connection and $SU(2)$ spin in hand, we can now write down a more explicit form of the action describing a single monopole interacting with an impurity. It is
\be S = \int dt\ \left(\frac{M}{2} \dot{X}^i\dot{X}^i + \frac{M}{2}\dot{\chi}^2 + \frac{\kappa_{\rm eff}}{2}\dot{\varphi}\cos\theta  + {\rm Tr}[JA_i(X)] \dot{X}^i + {\rm Tr}[J\phi(X)]\dot{\chi}\right)\ \ \ \ \ \label{msact}\ee
It's useful to examine each equation of motion in turn. The equation of motion governing the spin is
\be
\frac{dJ}{dt} = i [A_i \dot X^i + \varphi \dot{\chi} , J] 
\ee
This describes the precession of the spin in response to the motion of the monopole.

\para
From our discussion above, we know that as the spin precesses, its electric charge under $U(1)\subset SU(2)$ varies. This electric charge must be transferred to the monopole. 
Indeed, the  equation of motion for the dyonic degree of freedom $\chi$ reads
\be
M \ddot{\chi} +  {\Tr}[J B_i (X)]\dot{X}^i= 0
\label{claselec}\ee
Comparing to \eqn{mcharge} and \eqn{spincharge}, and making use of the Bogomolnyi equation \eqn{bog},  we see that this is simply the expression of the conservation of $U(1)$ charge $q_{\rm mono}+ q_{\rm spin}$.

\para
Finally, the equations of motion for the centre of mass degrees of freedom are
\be
M \ddot X^i = \epsilon^{ijk}\,{\rm Tr}[J B_k(X) ]\, \dot X_j + {\rm Tr}[J B_i (X)] \,\dot{\chi} 
\ee
The  right-hand side is simply the Lorentz force law. The first term is the velocity dependent force between the electrically charged impurity and the magnetic monopole; the second term is the Coulomb force that a dyon experiences in the presence of the impurity. Notice that the effective magnetic and electric fields experienced \emph{by} the monopole are the same as the  magnetic and electric fields \emph{of} the monopole. This, of course, is simply a manifestation of Newton's third law: the force that the spin exerts on the monopole is equal and opposite to the force that  the monopole exerts on the spin.

\section{Quantum Bound States}\label{boundsec}

In this section, we compute the quantum bound states of a monopole with a Wilson line. Since this is a discussion that is most natural in the context of supersymmetry, we start by describing the supersymmetric extension of our low-energy effective theory. 

\subsection{Supersymmetric  Dynamics}

The Yang-Mills action \eqn{4dact} can be extended to a theory with either ${\cal N}=2$ or ${\cal N}=4$ supersymmetry. Here we consider the ${\cal N}=2$ theory. In the absence of Wilson lines, monopoles are $1/2$-BPS, preserving ${\cal N}=(0,4)$ supersymmetry on their worldline. This means that the four collective coordinates $X^i$ and $\chi$ are joined by four real Grassmann collective coordinates $\xi^I$, $I=1,2,3,4$ that   can be interpreted as the Goldstino modes arising from the broken supersymmetries. 

\para
In the presence of a Wilson line, the monopoles remain $1/2$-BPS. There are no further Grassmann degrees of freedom associated to the Wilson line itself. Nonetheless, there are interesting ``spin-spin" interactions between the impurity and the Grassmann $\xi^I$.  To describe these, we first introduce some new notation.  We define 
\be A_I = (A_i,\phi)\ \ \  , \ \ \  X^I = (X^i, \chi)\ \ \ \ \ \ I=1,2,3,4\nn\ee
Then the Bogomolnyi equation \eqn{bog} becomes the self-dual Yang-Mills equation $F_{IJ}= {}^\star F_{IJ}$, supplemented with the requirement that $\partial_4=0$. In this notation, the action describing the interaction of the monopole and impurity \eqn{msact} can be written compactly as
\be S_{\rm mono-imp} = \int dt\  \left(\frac{M}{2} \dot{X}^I\dot{X}^I + \frac{\kappa_{\rm eff}}{2}\dot{\varphi}\cos\theta  + {\rm Tr}[JA_I(X)] \dot{X}^I\right)
\nn\ee
The  ${\cal N}=(0,4)$ supersymmetric completion of this action is 
\be
S_{\rm susy} =  S_{\rm mono-imp} + \int dt\ \left( i\frac{M}{2}  \xi^I \dot \xi^I   - \frac{i}{2}{\rm Tr}[J  F_{IJ} (X) ] \xi^I \xi^I \right) \label{susymact}\ee
The final term is the promised spin-spin interaction and will play an important role in determining the bound states. 

\para
It is not difficult to construct the four real supercharges. They are:
\be Q^i =  \frac M 2 \dot X^I \bar \eta_{IJ}^i \xi^J \ \ \ ,\ \ \  Q^4= \frac M 2 \dot X^I \xi^I
\nn\ee
 where $\bar{\eta}$ are the anti-self dual 't Hooft matrices. One can check that these are conserved for self-dual field strengths $F_{IJ}={}^\star F_{IJ}$. After canonical quantisation, they obey the ${\cal N}=4$ superalgebra  $\{Q_I, Q_J\} = \frac{1}{2} H\delta_{IJ}$.

\para
To see that we are dealing with an ${\cal N}=(0,4)$ algebra (as opposed to, say, ${\cal N}=(2,2)$), it is simplest to look at the R-symmetry of the theory which, in our case, is $SO(4)\cong SU(2)\times SU(2)$.   The  bosonic fields $X$ lie in the $({\bf 2},{\bf 1})$ representation while the fermions $\xi$ lie in the $({\bf 1},{\bf 2})$ representation.  The gauge connection $A_I$ also transforms as $({\bf 2},{\bf 1})$, but the fact that the field strength is self-dual means that it is a singlet under $SO(4)$; this is necessary in order that the spin-spin coupling $F_{IJ}\xi^I\xi^J$ is invariant. Finally, the four supercharges constructed above transform as $({\bf 2},{\bf 2})$ under the R-symmetry, as befits a theory with ${\cal N}=(0,4)$ supersymmetry.

\subsection{Quantum Bound States}

We now turn to the quantum mechanics of the supersymmetric theory \eqn{susymact}. The question that we would like to answer is: how many BPS bound states are there between a single monopole and a Wilson line? Such states were dubbed ``framed" BPS states in \cite{framed}. As we will see, even in this simple setting of a single monopole, the answer depends in an interesting manner on the representation of the Wilson line.

\subsubsection*{The Hilbert Space}

We begin by constructing the Hilbert space of the theory. Focussing initially on the impurity degrees of freedom, we have already seen in Section \ref{wilsec} that the Hilbert space is the appropriate representation of $SU(2)$, namely
\be |m\rangle\ \ \ \ \ \ m=-\frac{\kappa_{\rm eff}}{2},\ldots,\frac{\kappa_{\rm eff}}{2}\label{hilly}\ee
Usually we would refer to $\kappa_{\rm eff}/2$ as the total ``spin" of the representation, but we will be dealing with real (i.e. Lorentz) spins shortly so we shall avoid this terminology for $\kappa_{\rm eff}$.

\para
To construct the Hilbert space associated to the monopole degrees of freedom, we need to introduce the complex structure
\be z^1 = X^1 + iX^2 &\ ,\   z^2 = X^3 - i X^4 \ \ \ \ {\rm and}\ \ \ \ 
 \psi^1 = \xi^1 + i\xi^2\  ,\   \psi^2 = \xi^3 - i \xi^4 \nn\ee
 The Grassmann fields obey
 \be \{\psi^\sigma,\bar{\psi}^\rho\} = \frac{2}{M}\delta^{\sigma\rho}\ \ \ \ \sigma,\rho=1,2\nn\ee
 and can be used to build a four-dimensional Hilbert space, starting from a lowest weight state $|0\rangle$ obeying $\bar{\psi}^\sigma |0\rangle$, and building
 \be |0\rangle\ ,\ \psi^1|0\rangle\ ,\ \psi^2|0\rangle\ , \ \psi^1\psi^2|0\rangle\label{billy}\ee
 Of these, $|0\rangle$ and $\psi^1\psi^2|0\rangle$ are to be viewed as spin 0 monopoles. In contrast, $\psi^\sigma|0\rangle$ is a spin-$\frac{1}{2}$ monopole. 
 The full Hilbert space is constructed by  the tensor product of the two spaces \eqn{hilly} and \eqn{billy}, together with the spatial wavefunction for the monopole. The general state takes the form,
 \be |\Psi\rangle = \sum_{|m| \leq {\kappa_{\rm eff}}/{2} }\left( f_m (z, \bar{z}) + g_m (z,\bar{z} ) \, \psi^1 + h_m(z,\bar{z})\,\psi^2 + k_m(z,\bar{z}) \,\psi^1 \psi^2 \right) | m\rangle 
 \nn\ee
 where $\bar{\psi}^\sigma |m\rangle=0$ for $\sigma=1,2$ and each $m$. The gauge $SU(2)$ ``angular momentum" operators $J^a$ defined in \eqn{ja} act on this wavefunction as
\be
J^a\ \mapsto \ (T^a)^m_{\ n} 
\ee
where $(T^a)^m_{\ n}$, $a = 1,2,3$ are the generators of the $su(2)$ Lie algebra in the spin $\kappa_{\rm eff}/2$ representation.

\para
For our purposes, it will suffice to look at the action of single, complex supercharge, $Q$. We choose
\be
Q\ \mapsto\ \psi^\sigma\,D_{z^\sigma} \nn\ee
where $D_{z^\sigma} = \partial_{z^\sigma} - i A_{z^\sigma}^a T^a$. As before, one can check that the Hamiltonian is 
$H = \{ Q, Q^\dagger \}$

\subsubsection*{Conserved Charges}

\para
The monopole-impurity quantum mechanics has further conserved quantities. One of these is the $U(1)$ electric charge. We saw classically in \eqn{claselec} that this receives contributions from both the impurity and the dyonic degree of freedom of the monopole. Quantum mechanically, the electric charge operator is represented on the wavefunction as
\be
q  \mapsto\ - \frac i v \partial_{X^4}
\ee

\para
There is an important subtlety associated to this electric charge. We saw earlier that the corresponding collective coordinate $\chi = X^4$ has periodicity $2\pi/v$. But sending $X^4\rightarrow X^4+2\pi/v$ is equivalent to performing a gauge transformation $g=e^{2\pi i \phi/v}=-1$. This leaves the monopole invariant because it is built from adjoint valued fields. But, when $\kappa_{\rm eff}/2$ is half-integer, it flips the sign of the impurity degrees of freedom. This means that we should impose periodic or anti-periodic boundary conditions on the wavefunctions according to
\be
f_m (X^i, X^4 + 2\pi / v) = \left\{\begin{array}{cl} + f^m (X^i, X^4) \ \ \ \ \ & \kappa_{\rm eff} \in 2{\bf Z} \\ - f^m (X^i , X^4) \qquad & \kappa_{\rm eff} \in 2{\bf Z}+1 \end{array}\right.
\nn\ee
and similarly for $g_m$, $h_m$ and $k_m$. A similar requirement arises in the discussion of the dyon bound state spectrum in ${\cal N}=4$ super Yang-Mills  \cite{ashoke}.
As a consequence of these (anti)-periodic boundary conditions, the eigenvalues of $q$ are quantised in integer of half-integer multiples according to
\be
q\in 
 \left\{\begin{array}{cl} {\bf Z} \ \ \ \ \ & \kappa_{\rm eff} \in 2{\bf Z} \\ {\bf Z}+\frac{1}{2} \qquad & \kappa_{\rm eff} \in 2{\bf Z}+1 \end{array}\right.
 \nn\ee
This is to be expected: the dyon always carries integer electric charge while the impurity carries half-integer electric charge when $\kappa_{\rm eff}/2$ is half-integer.

\para
The remaining conserved quantities required for our discussion are the $SO(3)$ angular momenta associated with rotations in the $\mathbf R^3$ factor of the moduli space. They are represented quantum mechanically by the operators
\be
L_i \mapsto -i \epsilon_{ijk} \left( X^j \partial_{X^k} + \frac M 2 \xi^j \xi^k \right)  + T^i
\ee

\para
The conserved charges $H$, $q$, $L^2$ and $L_Z$ form a set of mutually commuting operators, implying that the energy eigenstates can be labelled by their electric charge and angular momentum quantum numbers.

\subsubsection*{Monopole-impurity bound states}

With this background, we now look for BPS bound states of our system. These are zero energy ground states, obeying 
\be
H \vert \Psi\rangle = 0 \quad \Leftrightarrow \quad Q \vert \Psi \rangle = Q^\dagger |\Psi\rangle = 0
\ee
These equations impose constraints on the wavefunctions $f_m$, $g_m$, $h_m$ and $k_m$. For the spin-0 wavefunctions $f_m$ and $k_m$, these constraints are simply that the functions are covariantly holomorphic,
\be
D_{z^1} f = D_{z^2} f = 0 \ \ \ {\rm and}\ \ \ D_{\bar z^1} k = D_{\bar{z}^2 } k = 0 
\nn\ee
There are no normalizable solutions to these equations. To see this, we can look at
\be
\int d^4X\ \left( \vert  D_{\bar{z}^1} f \vert^2 + \vert   D_{\bar{z}^2} f \vert^2 \right) & =&  \int d^4 X\  f^\dagger (-D_{z^1} D_{\bar{z}^1} -D_{z^2}  D_{\bar{z}^2} ) f \nn\ee
But, when written in complex coordinates, the self-duality condition $F_{IJ}= {}^\star F_{IJ}$ implies that $F_{z^1\bar{z}^1} + F_{z^2\bar{z}^2}=0$. This means that we can commute the covariant derivatives,
\be 
\int d^4 X\  f^\dagger (-D_{z^1} D_{\bar{z}^1} -D_{z^2}  D_{\bar{z}^2} ) f &=& \int d^4 X\ f^\dagger (- D_{\bar{z}^1}  D_{ z^1} - D_{\bar{z}^2} D_{ z^2} )f \nn\\ &=& 
  \int d^4X\  \left( \vert  D_{ z^1} f \vert^2 + \vert D_{ z^2} f \vert^2 \right)  = 0
\nn\ee
Hence $D_{\bar{z}^1} f =  D_{\bar{z}^2} f = 0$ as well. But this implies that $D_{X^I} D_{X^I} f = 0$, and $D_{X^I} D_{X^I}$ is a positive definite operator, so it is impossible for such a solution $f$ to exist.

\para
We have more joy with the spin-$\frac{1}{2}$ wavefunctions $g_m$ and $h_m$. The ground state equations mix these two functions together, requiring
\be
- D_{z^2} g + D_{z^1} h = 0\ \ \ {\rm and}\ \ \ \   D_{\bar{z}^1} g +  D_{\bar{z}^2} h = 0 \label{fghi}
\ee
These equations do possess a number of normalisable solutions, depending on the representation of the $SU(2)$ impurity determined by $\kappa_{\rm eff}$. To see this, we work in a fixed charge $q$ sector by writing
\be
g = \tilde g (X^i) e^{iqv X^4}\ \ \ ,\ \ \ h = \tilde h (X^i ) e^{iqv X^4}.
\nn\ee
and we define the two-component object
\be
\zeta (X^i) = \left( \begin{array}{c} + \tilde g (X^i) \\ -\tilde h (X^i) \end{array}\right)
\nn\ee
The equations \eqn{fghi} can then be written as a Dirac equation
\be
{\mathcal D} \zeta =0\label{clever}
\nn\ee
where the Dirac operator is 
\be
{\mathcal D} = \bar \sigma^I (\partial_I - i A_I^a  T^a) -qv  \qquad \quad  \bar \sigma^I = (\sigma^i , i1)
\nn\ee
Equations of this type have been studied in some detail in the literature. For $q=0$, this coincides with the equation for fermion zero modes in the background of a BPS monopole and was first studied in \cite{erick}. The equation has also been studied for $q\neq 0$ in the context of instanton zero modes in $d=2+1$ dimensional gauge theories \cite{ancient}; the instantons are again the BPS monopole solutions, while $qv$ plays the role of a real mass parameter. We now review the outcome of these computations.

\paragraph{}
To compute the  number of normalizable solutions to \eqn{clever}, we first introduce the regulated index \cite{erick}
\be
{\mathcal I} (\mu^2) = {\rm Tr} \left( \frac{\mu^2} {\mathcal D^\dagger \mathcal D + \mu^2} - \frac{\mu^2} { \mathcal D \mathcal D^\dagger + \mu^2} \right)\nn
\ee
The limit  ${\mathcal I} (\mu^2\rightarrow 0)$ counts the number of complex zero modes of $\mathcal D$ minus the number of complex zero modes of $\mathcal D^\dagger$. But  arguments similar to those sketched above show that $\mathcal D^\dagger$  is positive definite and has no zero modes, so the number of normalizable bound states is given by
\be
N =  \mathcal I (\mu^2\rightarrow 0)
\nn\ee
For $q\neq 0$, the index was evaluated in \cite{ancient}; the result is
\begin{equation}
\mathcal I (\mu^2) = \sum_{|m|\leq\kappa_{\rm eff}/2} \frac{m(m-q) v} { \sqrt{\mu^2 + (m-q)^2 v^2} }\label{before}
\end{equation}
From this, we deduce that the number of supersymmetric bound states of charge $q$ is
\be
N = \sum_{|m| \leq \kappa_{\rm eff}/2} m \,{\rm sign}( m- q-q\epsilon )
\label{after}\ee
Here $\epsilon > 0$ is a small number which is included to avoid counting marginally  non-normalisable states. To understand this, it is perhaps best to look at a simple example. Suppose we wish to count the number of normalisable bound states with a Wilson line in the fundamental representation. This corresponds to $\kappa_{\rm eff}=1$. In this case, the electric charge is necessarily half-integer. If we were to write $q=\frac{1}{2}+\epsilon$, the naive application of \eqn{before} gives $N=0$ bound states, while choosing $q=\frac{1}{2}-\epsilon$ gives $N=1$ bound states.  What's happening here is that as $q$ crosses the value $\frac{1}{2}$ from below, the bound state is becoming non-normalizable.  When $q$ is exactly $\frac{1}{2}$, this zero mode is marginally non-normalisable: the integral of the square of its wavefunction diverges, but only logarithmically. Since we do not wish to count marginally non-normalisable zero modes, we evaluate $N$ using $q = \frac{1}{2}+ \epsilon$.

%
%

\para
For $\vert q \vert \geq \kappa_{\rm eff}/2$, it is clear that $N = 0$ and there are no supersymmetric bound states.
For $\vert q \vert < \kappa_{\rm eff}/2$, we can perform the summation explicitly to obtain a formula for the number of supersymmetric bound states
\be
N = \left\{\begin{array}{ll} \ \ \frac{1}{4}\kappa_{\rm eff}(\kappa_{\rm eff}+2)  - q^2 - \vert q \vert & \kappa_{\rm eff} \in 2{\bf Z}\\ \ \  \frac{1}{4}(\kappa_{\rm eff} + 1)^2 - (q^2 + \frac{1}{4}) - \vert q \vert \ \ \ \ \ \ \ \ \ \ \ &  \kappa_{\rm eff}\in 2{\bf Z}+1 \end{array}\right.
\nn\ee

\para
The spectrum of BPS bound states can be decomposed into irreducible representations of the $SO(3)$ symmetry group of  spatial rotations in $\mathbf R^3$ . The action of the angular momentum operators on the two-component spinor $\zeta$ is given by
\be
L_i \zeta = \left(  -i \epsilon_{ijk} X^j \partial_{X^k} + \frac 1 2 \sigma^i + T^i \right) \zeta \nn
\ee

\para
These angular momentum operators commute with the Dirac operator $\mathcal D$, ensuring that the zero modes of  $\mathcal D$ decompose into $SO(3)$ multiplets. This decomposition was analysed in \cite{cox}, where it was shown that the zero mode spectrum of the Dirac operator contains at least one multiplet with angular momentum $l$ for each value of $l$ in the range
\be
l = \vert q \vert + \frac 1 2 \ , \ \vert q \vert + \frac 3 2 \ , \ ... \ \  , \frac {\kappa_{\rm eff}} 2 - \frac 1 2
\ee
This result, in conjunction with \eqn{after}, implies that there is exactly one multiplet of zero modes for each $l$ in this range.

\para
The table below summarises the spectrum of framed BPS states for Wilson lines in small representations of the gauge group.
\begin{center}
\begin{tabular}{c|ccccccccc}
$\kappa_{\rm eff}$&  $ q= -2 \ \ $ & $\ \  - \frac{3}{2} \ \ $ & $ \ \  - 1 \ \ $ & $ \ \ - \frac{1}{2} \ \ $ & $ \ \ 0 \ \ $ &  $ \ \ +\frac{1}{2} \ \ $ & $ \ \ +1 \ \ $ & $ \ \ +\frac{3}{2} \ \ $ & $ \ \ +2 \ \ \ $  \\
\hline
0  \\
$1$ \\
$2$ &&&&& $ \mathbf 2 $  \\
$3$ &&&& $ \mathbf 3 $ && $ \mathbf 3 $  \\
$4$ &&& $ \mathbf 4 $ && $ \mathbf 2 \oplus \mathbf 4$ && $ \mathbf 4 $  \\
$5$ && $\mathbf 5 $ && $ \mathbf 3 \oplus \mathbf 5$ && $ \mathbf 3 \oplus \mathbf 5$  && $ \mathbf 5 $  \\
$6$ & $ \ \ \  \ \mathbf 6 $ && $ \mathbf 4 \oplus \mathbf 6$ && $ \mathbf 2 \oplus \mathbf 4 \oplus \mathbf 6$ && $ \mathbf 4 \oplus \mathbf 6$ && $ \mathbf 6 $  \\
\end{tabular}
\end{center}

$ $

\section{Future Directions}

Supersymmetric gauge theories have been used as tractable, toy models to explore strongly coupled phenomena in high energy physics for many years. The idea that this can be extended to the kind of situations that may be relevant for condensed matter physics is an appealing one.  In the context of Abelian theories in $d=2+1$ dimensions, this  has been explored recently in \cite{shamit,shamit2} through the addition of electric and magnetic impurities which preserve (or, at the very least, only softly break) supersymmetry.

\para
For non-Abelian gauge theories, the situation seems somewhat richer. As we reviewed in Section \ref{wilsec}, the analog of a non-Abelian electric impurity is a Wilson line which can be represented as a localised spin degree of freedom. This raises the interesting possibility of an interplay between the dynamics of non-Abelian gauge theories and spin systems. For example, one could add a lattice of spins and study  the low-energy, continuum limit. In the ultra-violet, BPS spins exert no force on each other. However this could change, either through RG flow or through soft breaking of supersymmetry. It is clear that one can engineer situations in which the low-energy fields mediate interactions between spins, whether ferroelectric or anti-ferroelectric.  It  may be interesting to map out the possible phase structure consistent with supersymmetry.

\para
In Section \ref{monosec}, we presented a description of monopoles moving in the background of electric spins. This has applications to more mathematical aspects of supersymmetric gauge theories and provides a semi-classical method to compute a class of framed BPS states introduced in \cite{framed}. However, it can also be thought of more physically  as the dual to electrons moving in the background of magnetic spins and it is tempting to  think of this as a possible approach to a class of Kondo problems.

\section*{Acknowledgements}

We're grateful to Nick Dorey, Nick Manton, Andrew Singleton and David Skinner for very useful discussions and comments. We are supported by STFC and by the European Research Council under the European Union's Seventh Framework Programme (FP7/2007-2013), ERC Grant agreement STG 279943, Strongly Coupled Systems

\end{document}